\documentclass[prl,twocolumn,showpacs,amsmath,amssymb,floats,floatfix,prl,aps,superscriptaddress,showkeys]{revtex4}

\usepackage{graphicx}
\usepackage{epsfig}
\usepackage{dcolumn}
\usepackage{bm}
\usepackage{times}
\graphicspath{{C:/Users/Gabriel/Desktop/figures/}}
\begin{document}


\title{Invariance of Witten's quantum mechanics under point canonical transformations}

\author{Gabriel Gonz\'alez}

\email{gabriel.gonzalez@uaslp.mx}
\affiliation{C\'atedra CONCAYT, Universidad Aut\'onoma de San Luis Potos\'i, San Luis Potos\'i, 78000 MEXICO}
\affiliation{Coordinaci\'on para la Innovaci\'on y la Aplicaci\'on de la Ciencia y la Tecnolog\'ia, Universidad Aut\'onoma de San Luis Potos\'i,San Luis Potos\'i, 78000 MEXICO}

\date{\today}

\begin{abstract}
We show that the supersymmetric algebra of Witten's quantum mechanics is invariant under a given point canonical transformation. It is shown that Witten's supersymmetric quantum mechanics can be isospectral or not to the seed Hamiltonian depending on the space coordinate you work on. We illustrate our results by generating a new class of exactly solvable supersymmetric partner Hamiltonians which are not isospectral to the seed Hamiltonian.
\end{abstract}



\maketitle

\begin{figure}[ht!]
\includegraphics[width=8cm]{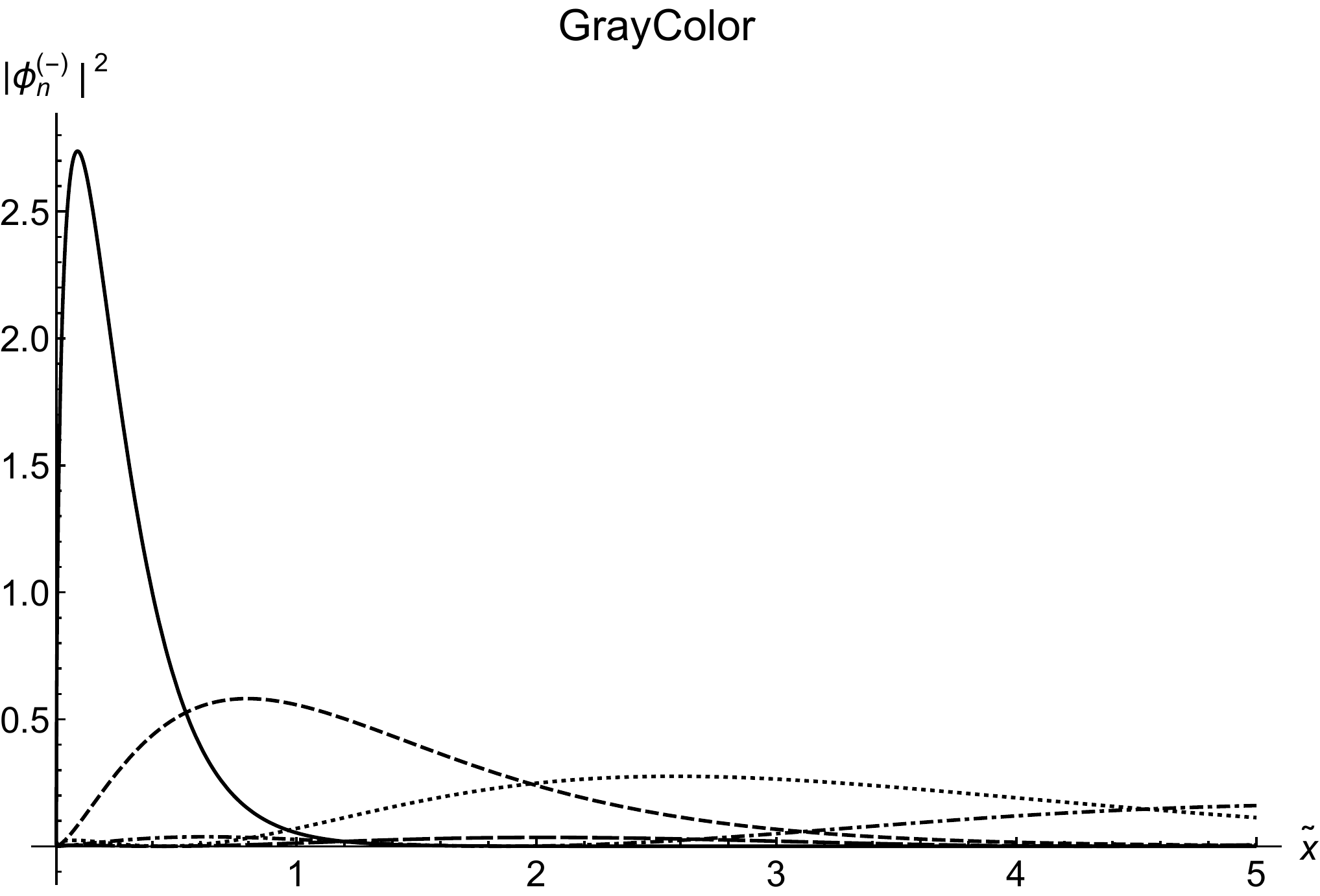}
\caption{The graph shows the probability distributions of the first stationary states of the partner Hamiltonian $\hat{H}_-$. Note how the $|\phi_n^{(-)}|^2$ goes to zero at the origin.}
\label{partner1}
\end{figure}
\begin{figure}[ht]
\includegraphics[width=8cm]{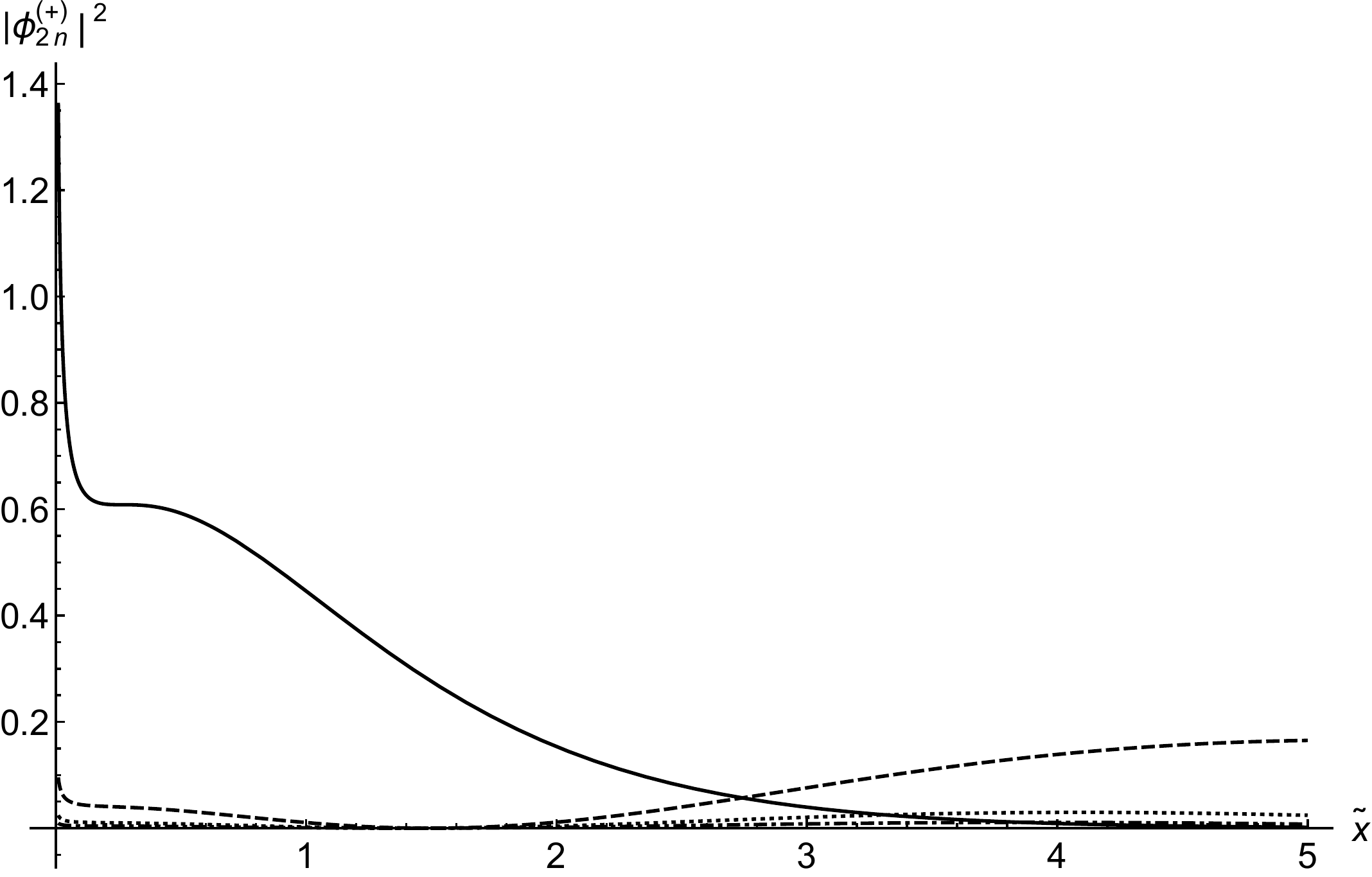}
    \caption{The graph shows the probability distributions for $n=0,2,4,6$ for the stationary states of the partner Hamiltonian $\hat{H}_+$. Note how the $|\phi_{2n}^{(+)}|^2$ blows up as ${\tilde x}\rightarrow 0$. Nevertheless, the eigenfunctions $\phi_{2n}^{(+)}$ are normalizable functions.}
	\label{partner2e}
\end{figure}
\begin{figure}[ht]
\includegraphics[width=8cm]{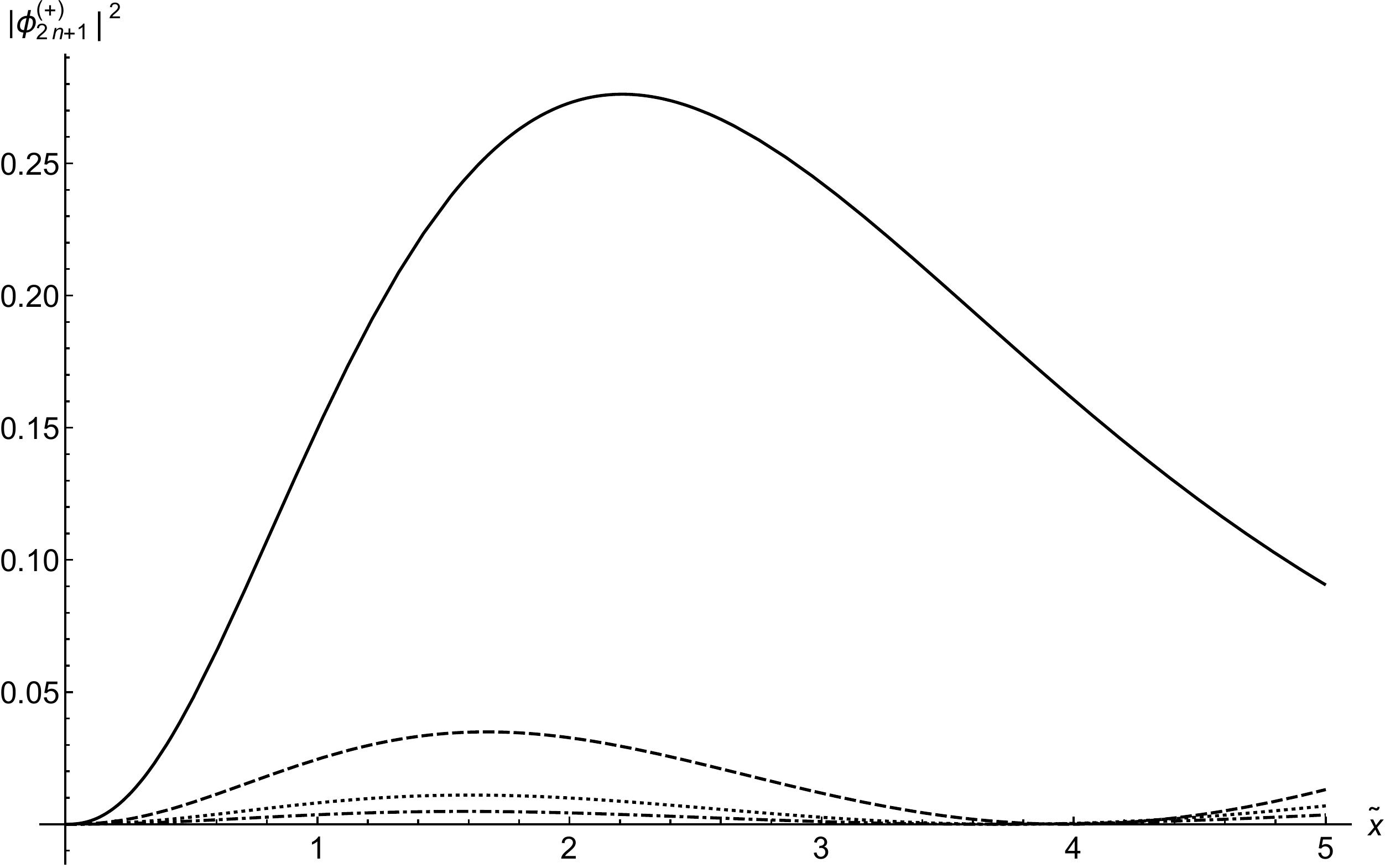}
    \caption{The graph shows the probability distributions for $n=1,3,5,7$ for the stationary states of the partner Hamiltonian $\hat{H}_+$. Note how the $|\phi_{2n+1}^{(+)}|^2$ goes to zero at the origin.}
	\label{partner2o}
\end{figure}

The solutions of the non relativistic Schr\"odinger equation are important because they often reveal the underlying symmetry and features of the physical system. The number of known exactly solvable quantum potentials has expanded considerably in recent years due to supersymmetry. The ideas of supersymmetry are based on the factorization method first proposed by Dirac in 1930 in order to solve the one dimensional harmonic oscillator.\cite{Dirac,schro,schro1} Years later, Infeld and Hull gave a classification of all systems solvable through the factorization method.\cite{Infeld,Infeld1,Infeld2} The generalization of the factorization method initiated with the work of Witten in 1981.\cite{Witten,Witten1} In Witten's one dimensional supersymmetric quantum mechanics the Hamiltonian of a given system is represented by a pair of supersymmetric partner Hamiltonians $H_{\pm}$ for which the energy of the ground state for $H_-$ is set to zero. Supersymmetry is said to be unbroken in the case when a zero energy state exists. Apart from the ground state of $H_-$, the energy levels of $H_{\pm}$ are isospectral. The energy states of the two partner Hamiltonians are related by first order differential operators through the so called intertwining relations. This procedure allow us to generate new solvable potentials starting from a known solution. In 1983 Gendenshtein showed that the eigenfunction and eigenvalues could be obtained algebraically whenever a parametric relation called shape invariant is satisfied by the two supersymmetric partner Hamiltonians.\cite{Genden} In 1984 Mielnik showed that there exists a family of equivalent partner Hamiltonians and obtained a family of quantum potentials isospectral to the harmonic oscillator.\cite{Mielnik, Fernandez, Adrianov1, Adrianov2,Adrianov3, Adrianov4} Later, Adrianov and Fern\'andez introduced second order differential intertwining operators by using two iterative first order transformations.\cite{Adrianov5, Adrianov6, Fernandez1}\\
There also exist other methods that yield new exactly solvable potentials that are not related with the factorization method and supersymmetry.\cite{Nieto, Sukumar, Sukumar1} The operator transformation procedure was used by Natazon who applied it to transform the Schr\"odinger equation into the hypergeometric and the confluent hypergeometric differential equation.\cite{Natanzon, Natanzon1, Negro} There are also other type of transformations called point canonical transformations (PCT) which transform a Schr\"odinger type equation into a known Schr\"odinger equation so that the solution is obtained through the known solution.\cite{De, Levay, Alhaidari, Quesne, Quesne1} Here we follow an alternative approach to obtain a new class of solvable potentials: starting from a well known exactly solvable Schr\"odinger equation we apply a point canonical transformation and determine the neccesary conditions that will leave Witten's supersymmetric quantum mechanics invariant. Therefore, the resulting transformation of coordinates will interrelate the Hilbert spaces and will give the energy spectrum and eigenfunctions in terms of those of the reference Hilbert space. The details of the method will be presented for the harmonic oscillator so that we will consider our reference Hilbert space spanned by the eigenfunctions of the harmonic oscillator. \\
Consider then the following time independent Schr\"odinger equation given by
\begin{equation}
\Big[\frac{d^2}{dx^2}-2m\left(U(x)-{\mathcal E}_n\right)\Big]\psi_n(x)=0,
\label{eq01}
\end{equation}
where we have set $\hbar=1$. If we apply to equation (\ref{eq01}) the following point canonical transformation
\begin{equation}
x=q({\tilde x}) \quad \mbox{and} \quad \psi_n(x)=\phi^{(-)}_n({\tilde x})g({\tilde x}),
\label{eq02}
\end{equation}
we get
\begin{widetext}
\begin{equation}
\Big[\frac{d^2}{d{\tilde x}^2}+\left(2\frac{g^{\prime}}{g}-\frac{q^{\prime\prime}}{q^{\prime}}\right)\frac{d}{d{\tilde x}}
+\left(\frac{g^{\prime\prime}}{g}-\frac{q^{\prime\prime}}{q^{\prime}}\frac{g^{\prime}}{g}\right)-2m(q^{\prime})^2\left(U(q({\tilde x}))-{\mathcal E}_n\right)\Big]\phi^{(-)}_n({\tilde x})=0.
\label{eq03}
\end{equation}
\end{widetext}
If we add and substract to equation (\ref{eq03}) the term $d[W({\tilde x})\phi^{(-)}_n({\tilde x})]/d{\tilde x}$ we have
\begin{widetext}
\begin{equation}
\frac{d}{d{\tilde x}}\Big[\frac{d\phi^{(-)}_n}{d{\tilde x}}+W\phi^{(-)}_n\Big]+\left(2\frac{g^{\prime}}{g}-\frac{q^{\prime\prime}}{q^{\prime}}-W\right)\frac{d\phi^{(-)}_n}{d{\tilde x}}
+\Big[\left(\frac{g^{\prime\prime}}{g}-\frac{q^{\prime\prime}}{q^{\prime}}\frac{g^{\prime}}{g}-\frac{dW}{d{\tilde x}}\right)-2m(q^{\prime})^2\left(U(q({\tilde x}))-{\mathcal E}_n\right)\Big]\phi^{(-)}_n=0.
\label{eq04}
\end{equation}
\end{widetext}
If we impose the following conditions over the point canonical transformation
\begin{align}
\label{eq05}
2\frac{g^{\prime}}{g} &=& \frac{q^{\prime\prime}}{q^{\prime}} \\ \nonumber
\frac{g^{\prime\prime}}{g}-\frac{q^{\prime\prime}g^{\prime}}{q^{\prime}g}-\frac{dW}{d{\tilde x}}-2m(q^{\prime})^2\left(U(q({\tilde x}))-{\mathcal E}_n\right)&=&E_n-W^2  
\end{align}
 where $E_n$ is an auxiliary constant, we can make the following substitution $d\phi^{(-)}_n/d{\tilde x}+W\phi^{(-)}_n=\sqrt{E_n}\phi^{(+)}_n$ into equation (\ref{eq04}) which gives us the following coupled differential equations
\begin{eqnarray}
\left(\frac{d}{d{\tilde x}}+W\right)\phi^{(-)}_n\!\!\!\!&=&\!\!\!\!\sqrt{E_n}\phi^{(+)}_n \rightarrow \hat{A}\phi^{(-)}_n\!\!=\!\!\sqrt{E_n}\phi^{(+)}_n\\ 
\left(-\frac{d}{d{\tilde x}}+W\right)\phi^{(+)}_n\!\!\!\!&=&\!\!\!\!\sqrt{E_n}\phi^{(-)}_n \rightarrow \hat{A}^{\dagger}\phi^{(+)}_n\!\!=\!\!\sqrt{E_n}\phi^{(-)}_n
\label{eq06}
\end{eqnarray}
The above equations are the so called intertwining relations and can be reduced to two uncoupled Schr\"odinger equations given by
\begin{eqnarray}
-\frac{d^2\phi^{(-)}_n}{d{\tilde x}^2}+\left(W^2-\frac{dW}{d{\tilde x}}\right)\phi^{(-)}_n=\hat{H}_-\phi^{(-)}_n=E_n\phi^{(-)}_n\\ 
-\frac{d^2\phi^{(+)}_n}{d{\tilde x}^2}+\left(W^2+\frac{dW}{d{\tilde x}}\right)\phi^{(+)}_n=\hat{H}_+\phi^{(+)}_n=E_n\phi^{(+)}_n.
\label{eq07}
\end{eqnarray}
Clearly, $W$ is the so called superpotential and $\hat{H}_{\pm}$ are supersymmetric partner Hamiltonians which can be factorized as $\hat{H_-}=\hat{A}^{\dagger}\hat{A}$ and $\hat{H_+}=\hat{A}\hat{A}^{\dagger}$, respectively.\\
Given the energy spectra ${\mathcal E}_n$ and eigenfunctions $\psi_{n}(x)$ of a physical system described by the potential energy function $U(x)$ we can choose a point canonical transformation $x=q(\tilde x)$ such that we can determine the superpotential and energy spectra of the supersymmetric partner Hamiltonians from equation (\ref{eq05}). To this end, we consider two possible point canonical transformations that will result in a constant term on the left hand side of equation (\ref{eq05}), which will be identified with the energy $E_n$. The first possibility is the identity coordinate transformation, i.e. $x=q({\tilde x})={\tilde x}$. For this choice of point canonical transformation we have $g({\tilde x})=q^{\prime}=1$ and substituting in equation (\ref{eq05}) we get
\begin{equation}
-\frac{dW}{d{\tilde x}}-2m\left(U(q({\tilde x}))-{\mathcal E}_n\right)=E_n-W^2
\label{eq08}
\end{equation}
Adding and subtracting $2m{\mathcal E}_0$ in equation (\ref{eq08}) we have
\begin{equation}
-\frac{dW}{d{\tilde x}}-2m\left((U(q({\tilde x}))-{\mathcal E}_0)-({\mathcal E}_n-{\mathcal E}_0)\right)=E_n-W^2
\label{eq09}
\end{equation}
Since the superpotential has to be independent of the index $n$ we have to choose $E_n=2m({\mathcal E}_n-{\mathcal E}_0)$. Note that we have set the energy of the ground state of $\hat{H}_-$ to zero. The energy spectra of the two partner Hamiltonians $\hat{H}_{\pm}$ are the same, the only exception is the ground state of $\hat{H}_-$, which lacks a counter part in $\hat{H}_+$.
The superpotential is obtained by solving the following Ricatti differential equation
\begin{equation}
-\frac{dW}{d{\tilde x}}-2mU_-(q({\tilde x}))=-W^2
\label{eq10}
\end{equation}
where $U_-(q({\tilde x}))=U(q({\tilde x}))-{\mathcal E}_0$. One can easily verify that a particular solution of equation (\ref{eq10}) is given in terms of the ground state wave function by
\begin{equation}
W({\tilde x})=-\frac{1}{\psi_0(q)}\frac{d\psi_0}{dq}q^{\prime}
\label{eq11}
\end{equation}
Therefore, the solutions of the two partner Hamiltonians with potentials $U_{\pm}(q({\tilde x}))=W^2 \pm dW/d{\tilde x}$ are related by $\phi^{(-)}_n({\tilde x})=\psi_n(x)$ and $\phi^{(+)}_n({\tilde x})=\hat{A}\phi^{(-)}_{n}({\tilde x})/\sqrt{E_{n}}$ for $n=0,1,2.\cdots$. Thus, the corresponding zero energy is missing from the spectrum of $\hat{H}_+$. \\
At this stage, it is natural to ask wether it is possible to find another point canonical transformation that will result in a constant term on the left hand side of equation (\ref{eq05}), which will be identified with the energy $E_n$. We find that this is indeed the case if we set 
\begin{equation}
\left(\frac{dq}{d{\tilde x}}\right)^2U(q)=\frac{m}{\sigma^2}
\label{eq12}
\end{equation}
where $\sigma^2$ is an adjustable parameter. If we substitute equation (\ref{eq12}) into equation (\ref{eq05}) we have
\begin{equation}
\frac{g^{\prime\prime}}{g}-\frac{q^{\prime\prime}g^{\prime}}{q^{\prime}g}-\frac{dW}{d{\tilde x}}-2m\left(\frac{m}{\sigma^2}-(q^{\prime})^2{\mathcal E}_n\right)=E_n-W^2
\label{eq13}
\end{equation} 
If we choose $W=g^{\prime}/g+\xi({\tilde x})$ in equation (\ref{eq13}) we have
\begin{equation}
-\frac{d\xi}{d{\tilde x}}-2m\left(\frac{m}{\sigma^2}-\frac{m{\mathcal E}_n}{\sigma^2U(q)}\right)=E_n-\frac{q^{\prime\prime}}{q^{\prime}}\xi-\xi^2
\label{eq14}
\end{equation}
In order to solve equation (\ref{eq14}) we need to specify first the potential $U(q)$ since the potential depends on the parameter $\sigma$. Therefore, we will solve equation (\ref{eq14}) for the case of the harmonic oscillator potential, i.e. $U(q)=m\omega^2q^2/2$. Using equation (\ref{eq12}) we have the following first order differential equation
\begin{equation}
\sqrt{U(q)}dq=\frac{\sqrt{m}}{\sigma}d{\tilde x}
\label{eq14a}
\end{equation}
From equation (\ref{eq14a}) we obtain $x=q(\tilde x)=\sqrt{2\sqrt{2}{\tilde x}/\omega\sigma}$. Note that this coordinate transformation is defined only for $0<{\tilde x}<\infty$. Using the solution of equation (\ref{eq14a}) we obtain for the potential energy $U(q({\tilde x}))=\sqrt{2}m\omega{\tilde x}/\sigma$ and the term $q^{\prime\prime}/q^{\prime}=-1/2{\tilde x}$. Substituting these two terms in equation (\ref{eq14}) we have
\begin{equation}
-\frac{d\xi}{d{\tilde x}}-2m\left(\frac{m}{\sigma^2}-\frac{{\mathcal E}_n}{\sqrt{2}\omega\sigma{\tilde x}}\right)=E_n+\frac{1}{2{\tilde x}}\xi-\xi^2
\label{eq14b}
\end{equation}
Since the superpotential has to be independent of the index $n$ we must choose $\sigma={\mathcal E}_n/m$.
Adding and subtracting $2m^4/{\mathcal E}_0^2$ into equation (\ref{eq14b}) we have
\begin{equation}
-\frac{d\xi}{d{\tilde x}}-2m\left(\frac{m^3}{{\mathcal E}_n^2}-\frac{m^3}{{\mathcal E}_0^2}+\frac{m^3}{{\mathcal E}_0^2}-\frac{m}{\sqrt{2}\omega{\tilde x}}\right)=E_n+\frac{1}{2{\tilde x}}\xi-\xi^2
\label{eq15}
\end{equation}
Choosing $E_n=2m^{4}({\mathcal E}_0^{-2}-{\mathcal E}_n^{-2})$ in equation (\ref{eq15}) we get the following Ricatti differential equation for $\xi$
\begin{equation}
-\frac{d\xi}{d{\tilde x}}-\frac{2m^4}{{\mathcal E}_0^{2}}+\frac{2m^2}{\sqrt{2}\omega{\tilde x}}=\frac{1}{2{\tilde x}}\xi-\xi^2
\label{eq16}
\end{equation}
One can easily verify that a particular solution to equation (\ref{eq16}) is given by $\xi({\tilde x})=2\sqrt{2}m^2/\omega$.
The general solution of equation (\ref{eq16}) might be obtained once we know a particular solution and will give us a family of superpotentials. If we let $\xi({\tilde x})=2\sqrt{2}m^2/\omega+1/z(\tilde x)$ and substitute in equation (\ref{eq16}) we get the following linear differential equation for $z$, i.e.
\begin{equation}
\frac{dz}{d{\tilde x}}=\left(\frac{1}{2{\tilde x}}-\frac{4\sqrt{2}m^2}{\omega}\right)z-1.
\label{eq16a}
\end{equation}
The general solution of equation (\ref{eq16a}) is given by
\begin{align}
z({\tilde x})&=&\frac{1}{2m}\sqrt{\frac{\omega\pi{\tilde x}}{\sqrt{2}}}e^{-4\sqrt{2}m^2{\tilde x}/\omega}Erfi\Big[\sqrt{\frac{4\sqrt{2}m^2{\tilde x}}{\omega}}\Big] \\ \nonumber 
&& + C\sqrt{{\tilde x}}e^{-4\sqrt{2}m^2{\tilde x}/\omega}
\label{eq16b}
\end{align}
where $Erfi[{\tilde x}]$ is the imaginary error function and $C$ is an integration constant.
We will restrict ourselves to use only the particular solution such that the superpotential for our case is given by $W({\tilde x})=q^{\prime\prime}/2q^{\prime}+\xi=-1/4{\tilde x}+2\sqrt{2}m^2/\omega$. \\
We are now ready to obtain the supersymmetric partner Hamiltonians for the harmonic oscillator with their energy spectra and eigenfunction using the standard rules of supersymmetric quantum mechanics. We will set $m=\omega=1$ for simplicity.
The results of using the point canonical transformation given by equation (\ref{eq14a}) are as follows:
\begin{equation}
E_n=8\left(1-\frac{1}{(2n+1)^2}\right)
\label{eq17}
\end{equation}
\begin{equation}
U_{-}({\tilde x})=8-\frac{3}{16{\tilde x}^2}-\frac{\sqrt{2}}{{\tilde x}}
\label{eq18}
\end{equation}
\begin{equation}
U_+({\tilde x})=8+\frac{5}{16{\tilde x}^2}-\frac{\sqrt{2}}{{\tilde x}}
\label{eq19}
\end{equation}
\begin{equation}
\phi_n^{(-)}({\tilde x})=\left(\frac{2^9\sqrt{2}{\tilde x}}{(2n+1)^5}\right)^{1/4}\frac{H_n\left(\sqrt{4\sqrt{2}{\tilde x}}\right)}{\sqrt{2^nn!\sqrt{\pi}}}e^{-2\sqrt{2}{\tilde x}/(2n+1)}
\label{eq20}
\end{equation}
\begin{equation}
\phi_{n}^{(+)}({\tilde x})=\frac{n+3/2}{2\sqrt{2((n+3/2)^2-1/4)}}\hat{A}\phi_{n+1}^{(-)},
\label{eq21}
\end{equation}
where $H_n(q)$ are the Hermite polynomial. Again, the corresponding zero energy is missing from the spectrum of $\hat{H}_+$. Interestingly, we see from equation (\ref{eq17}) that the energy spectra of the partner Hamiltonians $\hat{H}_{\pm}$ is not isospectral to the reference Hamiltonian, i.e. ${\mathcal E}_n=n+1/2$. \\
Figure (\ref{partner1}) shows the graph of $|\phi_n^{(-)}|^2$ as a function of ${\tilde x}$. We see that all the eigenfunctions for the partner Hamiltonian $\hat{H}_-$ are well behaved at the origin, i.e. $\lim\limits_{{\tilde x}\rightarrow 0}\phi_n^{(-)}\rightarrow 0$.\\ In figures (\ref{partner2e}) and (\ref{partner2o}) we show the graphs of $|\phi_n^{(-)}|^2$ as a function of ${\tilde x}$ for $n$ even and $n$ odd, respectively. From the figures we see that half of the eigenfunctions for the partner Hamiltonian $\hat{H}_+$ are well behaved at the origin, i.e. $\lim\limits_{{\tilde x}\rightarrow 0}\phi_{2n+1}^{(+)}\rightarrow 0$.\\
Note also that the supersymmetric partner Hamiltonian $\hat{H}_+$ has exactly the same structure as the radial equation.\cite{Koste} For this case, only the solutions given by $\phi_{2n+1}^{(+)}$ satisfy the correct boundary conditions for a Hydrogen like potential.

One can use this new approach to construct new superpotentials from energy dependent potentials. Energy dependent potentials for the non relativistic Schr\"odinger equation arise from momentum dependent interactions. \cite{Green,Green1} In our case we are going to obtain the superpotentials from a toy model of a harmonic oscillator under an energy dependent force. The energy dependent potential of this toy model is given by
\begin{equation}
U(q,{\mathcal E}_n)=\frac{1}{2}m\omega^2q^2-\lambda{\mathcal E}_n^{3/2}q=U_{HO}(q)-\lambda{\mathcal E}_n^{3/2}q
\label{eq21}
\end{equation} 
where $\lambda$ is a constant to ensure dimensional consistency. The unnormalized eigenfunctions of the Schr\"odinger equation for the potential given by (\ref{eq21}) are $\psi_n(x)=N_nH_n(\sqrt{m\omega}(x-\lambda{\mathcal E}_n/m\omega^2))e^{-m\omega(x-\lambda{\mathcal E}_n/m\omega^2)^2/2}$, with the energy spectra satisfying the following cubic equation:
\begin{equation}
{\mathcal E}_n^3+\frac{2m\omega^2}{\lambda^2}{\mathcal E}_n-\frac{2m\omega^3}{\lambda^2}\left(n+\frac{1}{2}\right)=0
\label{eq22}
\end{equation}
If we choose our point canonical transformation to be given by
\begin{equation}
\left(\frac{dq}{d{\tilde x}}\right)^2U_{HO}(q)=\frac{m}{\sigma^2}
\label{eq23}
\end{equation}
we get from equation (\ref{eq23}) te following point canonical transformation $x=q(\tilde x)=-\sqrt{2\sqrt{2}{\tilde x}/\omega\sigma}$. Note that this time we have chosen the coordinate transformation to be negative. Using the solution of equation (\ref{eq23}) we have the following equation for $\xi({\tilde x})$:
\begin{equation}
-\frac{d\xi}{d{\tilde x}}-2m\left(\frac{m^3}{{\mathcal E}_n^2}-\frac{m}{\sqrt{2}\omega{\tilde x}}\left(1-\lambda\sqrt{\frac{2\sqrt{2}m{\tilde x}}{\omega}}\right)\right)=E_n+\frac{1}{2{\tilde x}}\xi-\xi^2
\label{eq24}
\end{equation}
where we have chosen $\sigma={\mathcal E}_n/m$.
Adding and subtracting $2m^4/{\mathcal E}_0^2$ into equation (\ref{eq24}) and choosing $E_n=2m^{4}({\mathcal E}_0^{-2}-{\mathcal E}_n^{-2})$ we get the following Ricatti differential equation 
\begin{equation}
-\frac{d\xi}{d{\tilde x}}-\frac{2m^4}{{\mathcal E}_0^{2}}+\frac{2m^2}{\sqrt{2}\omega{\tilde x}}\left(1-\lambda\sqrt{\frac{2\sqrt{2}m{\tilde x}}{\omega}}\right)=\frac{1}{2{\tilde x}}\xi-\xi^2
\label{eq25}
\end{equation}
One can easily verify that a particular solution to equation (\ref{eq25}) is given by 
\begin{equation}
\xi({\tilde x})=\frac{\sqrt{2}m^2}{{\mathcal E}_0}+\frac{\lambda{\mathcal E}_0}{2\omega^{3/2}}\sqrt{\frac{2\sqrt{2}m}{{\tilde x}}}.
\label{eq26}
\end{equation}
The superpotential for this case is given then by
\begin{equation}
W({\tilde x})=-\frac{1}{4{\tilde x}}+\frac{\sqrt{2}m^2}{{\mathcal E}_0}+\frac{\lambda{\mathcal E}}{2\omega^{3/2}}\sqrt{\frac{2\sqrt{2}m}{{\tilde x}}}.
\label{eq27}
\end{equation}
The supersymmetric partner potentials for this case are given by
\begin{equation}
U_{-}({\tilde x})=\frac{2}{{\mathcal E}_0^2}-\frac{3}{16{\tilde x}^2}+\frac{{\mathcal E}_0^3\lambda^2-1}{\sqrt{2}{\mathcal E}_0{\tilde x}}+2\lambda\sqrt{\frac{\sqrt{2}}{{\tilde x}}}
\label{eq28}
\end{equation}
\begin{equation}
U_+({\tilde x})=\left(\frac{\sqrt{2}}{{\mathcal E}_0}+\frac{\lambda{\mathcal E}_0}{\sqrt{\sqrt{2}{\tilde x}}}-\frac{1}{4{\tilde x}}\right)^2+\frac{1}{4{\tilde x}^2}-\frac{\lambda{\mathcal E}_0}{4}\left(\frac{\sqrt{2}}{{\tilde x}}\right)^{3/2}.
\label{eq29}
\end{equation}
where we have set $m=\omega=1$. Note that the partner superpotentials given in equations (\ref{eq28}) and (\ref{eq29}) reduce to the partner superpotentials given by equations (\ref{eq18}) and (\ref{eq19}) when $\lambda\rightarrow 0$ and ${\mathcal E}_0\rightarrow 1/2$.\\

In conclusion, we have demonstrated that for a given potential $U(x)$ and energy spectra ${\mathcal E}_n$ the algebra of unbroken supersymmetric quantum mechanics can be obtained through a coordinate transformation given by $x=q({\tilde x})$ where $q({\tilde x})=S^{-1}(\sqrt{m}{\tilde x}/\sigma)$ where $S(y)=\int{\sqrt{U(y)}}dy$. \\This coordinate transformation is used to generate a new class of supersymmetric partner Hamiltonians which are not isospectral to the seed Hamiltonian, in contrast to standard supersymmetric quantum mechanics. These results provide a new way to obtain exactly solvable potentials in one dimensional quantum mechanics. \\
We should point out that there might be other point canonical transformations that could result in a constant term in the left hand side of equation (\ref{eq05}) that could be interpreted as $E_n$ and that will result in other classes of supersymmetric partner Hamiltonians.\\ \\
I would like to acknowledge support by the program ``C\'atedras CONACYT" through project 1757 and from project 105 of ``Centro Mexicano de Innovaci\'on en Energ\'ia Solar" and by the National Laboratory program from CONACYT through the Terahertz Science and Technology National Lab (LANCYTT).

\end{document}